\documentclass[12pt]{article}

\def\be{\begin{equation}}
\def\ee{\end{equation}}
\def\a{\alpha}

\def\s{\sigma}
\def\G{\Gamma}
\def\pd{\partial} 
\def\l{\lambda} 
\def\det{\mbox{det}}
\def\ch{\mbox{ch}}

\def\ln{\mbox{ln}}
\def\exp{\mbox{exp}}
\def\cos{\mbox{cos}}
\def\sin{\mbox{sin}}
\def\ctg{\mbox{ctg}}
\def\tg{\mbox{tg}}
\def\ak{a^{+}}
\def\ck{c^{+}}
\def\ra{\rangle}
\def\la{\langle}
\def\D{\Delta}
\def\n{\tilde{n}}
\def\r{\rho}

\begin{document}

\begin{center}
{\bf Formfactors and functional form of correlators in the XX-spin chain. }
\end{center}
\vspace{0.2in}
\begin{center}
{\large A.A.Ovchinnikov}
\end{center}   
\begin{center}
{\it Institute for Nuclear Research, RAS, 117312, Moscow}
\end{center}   

\vspace{0.2in}

\begin{abstract}

We present the new expressions for the formfactors of local operators 
for the XX - quantum spin chain as a Cauchy determinants. Using the known 
functional form of the correlator at large distances we propose 
the new expression for the constant for the asymptotics of the correlator 
as a Cauchy determinant. 
We calculate the momentum distribution for the general case of 
the XXZ - spin chain and point out that it is completely different from 
that for the Luttinger model (the system of fermions). 
For the XX chain we compare numerically the value of the lowest formfactor 
and the expectation value of momentum- zero operators which is determined 
by the functional form of the correlator.

\end{abstract}

\vspace{0.1in}

{\bf 1.Introduction.}

In the present paper we study the spin-spin equal-time correlator in the XX - spin 
chain with the Hamiltonian  
$H=\frac{1}{2}\sum_{i=1}^L(\sigma^x_i\sigma^x_{i+1}+\sigma^y_i\sigma^y_{i+1})$, 
which is the particular case of the XXZ spin chain with the anisotropy 
parameter $\Delta=0$.  
It is worth mentioning that although the XX-chain can be solved with the help 
of mapping to the free-fermion system via the Jordan-Wigner transformation the 
system is essentially the the hard -core bosons on the lattice (at the half-filling) 
and is not equivalent to the free fermions which manifests for example in the 
completely different correlators in the two models. 
      The interest in the calculation of correlators for the XX spin chain 
is in particular due to the possibility to use it as a testing ground for 
various approaches for the calculation of the correlation functions for 
the general case of the XXZ quantum spin chain and the other models 
solvable with the help of the algebraic Bethe ansatz method.   
    The correlators for the XX spin chain as well as the correlators for the Ising 
- type spin chains both at zero and finite temperature where previously studied 
long time ago \cite{LSM}, \cite{McCoy}, \cite{Wu}, \cite{Barouch} using various 
methods based on the application of theorems for the Toeplitz determinants. 
Recently it was also shown \cite{O} that  
the equal-time spin-spin correlator in the XX- model can 
be calculated exactly on a chain of finite length at any distance $x$ of order 
of the chain length and  the functional 
form of the correlator coincides with that predicted by the 
conformal field theory \cite{Cardy} or the bosonization procedure 
\cite{ML}, \cite{M}, \cite{LP}.

The goal of the present paper is twofold. First, we study the consequences 
of the functional form of the correlator for the XX chain in particular 
for estimates of the certain formfactors of local operators for this model. 
The interest in the formfactors is in particular, due to the existence of the 
closed expressions for the formfactors for the XXZ model in the form of the 
determinants obtained in the framework of the algebraic Bethe ansatz method  
(for example, see \cite{KMT}).  
Here we present the new expressions for the formfactors of local operators 
for the XX - quantum spin chain as a Cauchy determinants. Using the known 
functional form of the correlator at large distances we propose 
the new expression for the constant for the asymptotics of the correlator 
as a Cauchy determinant or in the form of the product depending on the 
momenta. At the same time using the simple scaling arguments the numerical 
value of the ``lowest'' formfactor is found. This value is compared with 
the values of the other formfactors, which shows their drastic decrease  
with the increasing of the energy or momentum of the intermediate state. 
We discuss the consequences of this fact. 
The second goal is to calculate the momentum distribution for the general 
case of the XXZ - spin chain and point out that it is completely different 
from that for the Luttinger model (the system of fermions). 
The spectral properties (momentum distribution) are of interest for both  
models due to their possible realizations in nature. 
In particular we point out the universal character of the connection of 
the momentum distribution singularity with the asymptotics of the correlators 
and the constants in front of the asymptotics.  
For the Luttinger model the connection of the momentum distribution with the 
functional form of the correlators is obscure in the literature.  
The momentum distribution for the XXZ spin chain 
have not been considered previously.

In Section 1 we show how the usual bosonization procedure leads to the 
functional form predicted by the conformal invariance and discuss  
the momentum distribution for the system of spinless fermions 
(Luttinger model).  
In Section 2 we calculate the formfactors and obtain the new expression 
for the formfactors in the form of the Cauchy determinant.  
In Section 3 we briefly review the well known calculation of equal-time 
spin-spin correlator in the XX -model \cite{LSM}, \cite{McCoy} and 
present the expression for the correlator at the distances $x\sim L$. 
We then use the functional form and the expressions for the formfactors 
to get the new expression for the constant in front of the asymptotics. 
Later in this section we evaluate the momentum distribution for the XXZ 
spin chain. 
In the Appendix we present for completeness
some formulas and the numerical constants used in the 
text and also briefly review the previous calculation of the constant 
in front of the asymptotics for the XX spin chain.

\vspace{0.2in}

{\bf 1. Bosonization.}

\vspace{0.1in}

Consider the effective low -energy Hamiltonian which build up from 
the fermionic operators ($a_k,~c_k,~k=2\pi n/L,~n\in Z$, $L$- is the length 
of the chain) corresponding to the excitations around the right and the 
left Fermi- points and consists of the kinetic energy term and the 
interaction term $H=T+V$ with the coupling constant $\l$:    
\be
H=\sum_kk(a_k^{+}a_k-c_k^{+}c_k)+\l/L\sum_{k,k',q}\ak_ka_{k+q}\ck_{k'}c_{k'-q}.      
\label{ham}
\ee
Defining the operators \cite{ML}
\[
\r_1(p)=\sum_k\ak_{k+p}a_k,~~~~\r_2(p)=\sum_k\ck_{k+p}c_k,
\]
where $|k|,|k+p|<\Lambda$, where $\Lambda$ is some cut-off energy, 
which for the states with the filled Dirac sea have the following 
commutational relations 
\[
\left[\r_1(-p);\r_1(p')\right]=\frac{pL}{2\pi}\delta_{p,p'}~~~~
\left[\r_2(p);\r_2(-p')\right]=\frac{pL}{2\pi}\delta_{p,p'},
\]
one can represent the Hamiltonian in the following form: 
\[
H=\frac{2\pi}{L}\sum_{p>0}\left(\r_1(p)\r_1(-p)+\r_2(-p)\r_2(p)\right)+ 
\l\sum_{p>0}\frac{2\pi}{L}\left(\r_1(p)\r_2(-p)+\r_1(-p)\r_2(p)\right).
\]
To evaluate the correlators in the system of finite length and make the 
connection with the conformal field theory predictions, one can introduce 
the lattice fields $n_{1,2}(x)$ corresponding to the Fourier transform of the 
operators $\r_{1,2}(p)$ and 
represent the last Hamiltonian 
in the sector with the total number of particles and the momentum 
$\D N=\D N_1+\D N_2,~~\D Q=\D N_1-\D N_2$, 
where $\D N_{1,2}$ are the numbers of particles at the two  
Fermi-points,  
in the following form (this was 
first proposed in ref.\cite{Haldane}): 
\be
H= 2\pi\sum_x\left(\frac{1}{2}(n_1^2(x)+n_2^2(x))+\lambda n_1(x)n_2(x)\right)
+\frac{\pi}{2L}u(\l)
\left[\xi(\D N)^2+(1/\xi)(\D Q)^2\right],
\label{finite}
\ee
where the parameters $u(\l)=(1-\l^2)^{1/2}$ and $\xi=((1+\l)/(1-\l))^{1/2}$. 
Calculation of finite - size corrections to the energy of the ground state for 
the XXZ- spin chain (see for example \cite{Kar}) leads to the expression  
(\ref{finite}) and allows one to obtain the parameter $\xi$ which leads to 
the predictions of critical indices according to the conformal field theory. 
The calculation gives the value $\xi=2(\pi-\eta)/\pi$, where the parameter $\eta$ 
is connected with the anisotropy parameter of the XXZ - chain as 
$\Delta=\cos(\eta)$ (in the present paper we consider the antiferromagnetic XXZ 
chain in the massless regime $0<\Delta<1$).  
Rescaling the variables as 
$\n_{1,2}(x)=\sqrt{2\pi}~ n_{1,2}(x)$
one finds the commutational relations
$\left[\n_1(x);\n_1(y)\right]=\delta^{\prime}(x-y)$ in the 
continuum limit which allow one to introduce    
following conjugated field and the momenta:
\[
\pi(x)= \frac{1}{\sqrt{2}}(\n_1(x)-\n_2(x));~~~~  
\pd_x\phi(x)= \frac{1}{\sqrt{2}}(\n_1(x)+\n_2(x))
\]
\[
\phi(x)= \tilde{N}(x)=\tilde{N}_1(x)+\tilde{N}_2(x),~~~ 
\tilde{N}_{1,2}(x)=\int_0^x dy~\n_{1,2}(y) 
\]
In terms of these variables the Hamiltonian density 
takes the following form: 
\be
H=\frac{1}{2}u(\l)\left[ (1/\xi)\pi^2(x)+ \xi(\pd\phi(x))^2\right]
 = \frac{1}{2}u(\l)\left[ \hat{\pi}^2(x)+ (\pd\hat{\phi}(x))^2\right], 
\label{h}
\ee
where 
\be
\pi(x)=\sqrt{\xi}~\hat{\pi}(x),~~~ 
\phi(x)=(1/\sqrt{\xi})\hat{\phi}(x).  
\label{canon}
\ee
The last equation (\ref{canon}) is nothing else but the canonical transformation,  
which is equivalent to the Bogoliubov transformation for the original operators 
$\rho_{1,2}(p)$. Next to establish the expressions for Fermions one should use 
the commutational relations $\left[a^{+}(x);\rho_{1}(p)\right]=-e^{ipx}a^{+}(x)$
and the same for $c^{+}(x)$. Note that these last relations were obtained using the 
expression with original lattice fermions: 
$\rho_{1}(p)=\sum_y e^{ipy}a^{+}(y)a(y)$. 
In this way we obtain the following expressions for fermionic operators:
\be
a^{+}(c^{+})(x)=
K_{1,2}~\exp\left(\pm\frac{2\pi}{L}\sum_{p\neq0}\frac{\rho_{1,2}(p)}{p}e^{-ipx}
\right)=K_{1,2}~\exp\left(\mp i2\pi N_{1,2}(x)\right),
\label{fermions}
\ee
where the fields $N_{1,2}(x)$ differ by the normalization $\sqrt{2\pi}$ from the 
fields $\tilde{N}_{1,2}(x)$ and $K_{1,2}$ are the Klein factors - the operators 
which creates the single particle at the right (left) Fermi -points  
(we omit here the usual exponential suppression $e^{-\a|p|/2}$ in the exponent and 
and the constant factor in front of the exponent $1/\sqrt{2\pi\a}$ which in the 
limit $\a\rightarrow 0$ leads to the correct anticommutational relations \cite{LP}). 
Note that the above expressions (\ref{fermions}) are equivalent to the known
``field-theoretical'' bosonization formulas \cite{M}.  
 
Now let us apply the above formulas to the specific case of the XXZ - spin chain. 
Using the Jordan-Wigner transformation $\sigma^{+}_x=a^{+}_x\exp(i\pi N(x))$,
where $a^{+}_x$ stands for the ``original'' lattice fermionic operator,  
and performing the obvious substitutions $N(x)\rightarrow x/2+N_1(x)+N_2(x)$ and
$a^{+}_x\rightarrow e^{ip_Fx}a^{+}(x)+e^{-ip_Fx}c^{+}(x),~~p_F=\pi/2$,  
we obtain after the canonical transformation (\ref{canon})
the expression for the spin operator which determines 
the leading term in the asympotics of correlator for the XXZ -chain:
\be
\sigma^{+}_x\sim(-1)^x\exp\left(-i\pi\sqrt{\xi}(\hat{N}_1(x)-\hat{N}_2(x))\right),
\label{leading}
\ee
where $\hat{N}_{1,2}(x)$ - are corresponds to the free fields 
$\hat{\pi}(x)$, $\hat{\phi}(x)$, obtained after the transformation (\ref{canon}). 
To these operators correspond the new operators $\rho_{1,2}(p)$ and the new 
fermionic operators (quasiparticles). 
Analogously the term responsible for the subleading asymptotics has the form 
\[
\exp\left(i2\pi(1/\sqrt{\xi})(\hat{N}_1(x)+\hat{N}_2(x))\right)
\exp\left(-i\pi\sqrt{\xi}(\hat{N}_1(x)-\hat{N}_2(x))\right). 
\]
Averaging the product of exponents in bosonic operators 
for the expression (\ref{leading}) 
and using the properties of $\rho_{1,2}(p)$, 
$\la\rho_1(-p)\rho_1(p)\ra=\frac{pL}{2\pi}\theta(p)$ and 
$\la\rho_2(p)\rho_2(-p)\ra=\frac{pL}{2\pi}\theta(p)$, 
we get for the correlation function 
$G(x)=\la0|\sigma^{+}_{i+x}\sigma^{-}_{i}|0\ra$ the following sum in 
the exponent: 
\[
C~\exp\left(~\frac{\xi}{4}~\sum_{n=1}^{\infty}\frac{1}{n}e^{in(2\pi x/L)}
+ h.c.\right), 
\]
where $C$ - is some constant. Then using the formula   
$\sum_{n=1}^{\infty}\frac{1}{n}z^n=-\ln(1-z)$ and substituting the value
$\xi=2(\pi-\eta)/\pi\rightarrow 1$ we obtain the following 
expression for the XX - chain: 
\be
G(x)=C_0\frac{(-1)^x}{\left(L\sin(\frac{\pi x}{L})\right)^{\alpha}},
~~~\alpha=\frac{\xi}{2}=\frac{\pi-\eta}{\pi}=1/2 ~~~(x>>1).
\label{coas}
\ee
Thus, although bosonization, which deals with the low-energy effective theory, 
is not able to predict the constant before the asymptotics, the critical exponent 
and the functional form are predicted in accordance with conformal field theory.

Let us show how the general 
form of the correlators, predicted by CFT, including the higher order terms 
in the asymptotics and their functional form can be obtained in the framework 
of bosonization for the XX chain and the spinless fermion model.  
For the XX chain 
we use the same formula for the operator $\sigma^{+}_x=a^{+}_x\exp(i\pi N(x))$.  
Next, the lattice 
fermionic operator $a^{+}_x$ should be projected on to the effective continuous 
operators $a^{+}(x)$, $c^{+}(x)$ which enter the effective low-energy theory 
(Luttinger model). Clearly, each of the operators should be accompanied by 
the combinations of the type $(c^{+}(x)a(x))^{m}$, $(a^{+}(x)c(x))^{m}$ 
conserving the total number of particles. 
Then one should consider the following substitution for the 
lattice operator 
\[
a^{+}_x=\sum_{m}\left(e^{ip_Fx+i2p_Fmx}C_1(m)a^{+}(c^{+}a)^{m}(x)+
e^{-ip_Fx-i2p_Fmx}C_2(m)c^{+}(a^{+}c)^{m}(x)\right), 
\]
where $p_F=\pi/2$ and  $C_1(m)$, $C_2(m)$  are some constants for an 
arbitrary integer number $m$. 
Clearly, repeating the procedure leading to the expression (\ref{coas}) 
we will obtain the expression for the correlator as a sum of the terms of the 
form (\ref{coas}) with the critical exponents $\a(m)=\a+m^2/\a$ in agreement with 
predictions of the conformal field theory. Note that we considered only the 
operators which conserve the number of particles, in general we would obtain 
the spectrum of primary operators labeled by two quantum numbers $n$ and $m$ of 
the form $\a(n,m)=n^{2}\a+m^2/\a$ in agreement with the prediction of the 
Gaussian model of the conformal field theory ($\sqrt{\xi}$ plays the role of 
the compactification radius).  
 In general there can be the  contributions 
corresponding to the descendant operators, which amounts to the presence of 
the additional terms of the type $(\pd_{x})^{k}N_{1,2}(x)$ in the last equation 
for $a^{+}_x$, however, one can argue that these terms are absent in the case 
of the XX spin chain.        
Let us stress that the correlator $G(x)$ is represented as a sum of the terms,  
each of the functional form (\ref{coas}) with the corresponding critical 
exponents $\a(m)=\a+m^2/\a$.  
Note that in the similar way the asymptotics and the functional form of 
each term in the expansion of the density-density correlation function 
can be found. For this case the coefficient for the leading term is equal 
to unity.

 In Section 3 we will compute the momentum distribution 
$\la n_k\ra=\la\s_k^{+}\s^{-}_k\ra$,  
which is connected with the asymptotics and the functional 
form of the correlator $G(x)$ and the behaviour of the formfactors at small 
energies or momenta $q$. Here before studying the momentum distribution 
and their relation to the formfactors in the XX spin chain, 
let us comment on the momentum distribution  
for the system of fermions, or for the Luttinger model (\ref{ham}). 
As an underlying model one can consider, for example, the exactly solvable 
system of fermions with the XXZ - type interaction, 
$H=\sum_{i}(-\frac{1}{2}(a_i^{+}a_{i+1}+h.c.)+\Delta n_{i}n_{i+1})$, 
for which the parameters of the Luttinger model and the critical exponents 
can also be calculated exactly. For this model the bosonization procedure 
leads to the following asymptotic expression for the correlator 
$G_F(x)=\la a_x^{+}a_0\ra$ analogous to the correlator (\ref{coas}):  
\be 
G_F(x)=C(p_F,\Delta) 
\frac{\sin(p_{F}x)}{\left(L\sin(\frac{\pi x}{L})\right)^{\alpha(\l)}},
~~~~~(x>>1),  
\label{sinas}
\ee
where $p_F$ is an (arbitrary) Fermi momenta, $C$ is some constant depending on 
$p_F$ and the other parameters of the model ($\Delta$) and the critical 
exponent equals 
\[
\a(\l)=
\frac{1}{2}\left(\xi+\frac{1}{\xi}\right)=\frac{1}{\sqrt{1-\l^2}} > 1.  
\]
Clearly at $\Delta=0$ we have $\l=0$, $\a(\l)=1$ and the correlator 
(\ref{sinas}) reduces to the free-fermion correlator. 
One can see from (\ref{sinas}) that the form of the correlator 
and the critical exponents in the two models are completely different 
although the thermodynamic quantities are the same.  
Calculating the Fourier transform of the correlator (\ref{sinas}) taking  
into account its functional form, one readily obtains the well known 
momentum distribution for the Luttinger model close to the Fermi points:   
\be
\la n_k\ra= \la a_k^{+}a_k\ra= \frac{1}{2}+C |k-p_F|^{\gamma}, ~~~~
\gamma=\a(\l)-1=(\xi^{1/2}-1/\xi^{1/2})^2/2.   
\label{nkf}
\ee
First, both this expression and the critical exponent $\gamma$ are exact 
in a sense that they are not an artifact of approximation of the initial 
lattice spinless fermion model by the Luttinger model. 
Both the function (\ref{nkf}) and the value of $\gamma$ are the direct 
consequence of the equation (\ref{sinas}) valid in the initial model 
so that $\gamma$ is directly connected with $\a(\l)$ in eq.(\ref{sinas}).  
Second, the constant $C$ in the equation (\ref{nkf}) related in a simple way 
with the constant $C(p_F,\Delta)$ in eq.(\ref{sinas}), 
the fact that have not been pointed out previously. 
In fact, the subleading terms in the correlator lead to the contributions 
to (\ref{nkf}) which are of higher order in $|k-p_F|$ and can be neglected 
in the vicinity of the Fermi point. Since eq.(\ref{nkf}) was obtained 
from (\ref{sinas}) for $k\to p_F$, the constant $C$ is proportional to 
$C(p_F,\Delta)$ with the coefficient which depends only on the exponent 
$\a(\l)$ in exactly the same way as for the constant in front of the 
asymptotics of $G(x)$ for the XXZ spin chain. 
We will derive this relation below in Section 3.   
Although the behaviour of the momentum distribution for the XXZ spin chain 
which is determined by the asymptotic behaviour (\ref{coas}) 
is completely different from that in the Luttinger model, the leading 
order singularity is also unambiguously predicted by the exponent $\a$ 
and the constant before the singularity is 
related in a simple way to the constant $C_0$.

\vspace{0.2in}

{\bf 2. Formfactors for the XX -spin chain.}

\vspace{0.1in}

Let us evaluate the expressions for the formfactors of the spin operators
$\sigma_i^{\pm}$ for the XX-model at zero magnetic field with the Hamiltonian:
\[
H=-\frac{1}{2}\sum_{i=1}^L(\sigma^x_i\sigma^x_{i+1}+\sigma^y_i\sigma^y_{i+1}), 
\]
where the periodic boundary conditions are implied. 
Performing the well-known Jordan-Wigner transformation:
\[
\sigma^{+}_{x}=e^{i\pi N(x)}a^{+}_{x}=\exp(i\pi\sum_{l<x}n_l)a^{+}_{x},
\]
we obtain the following Hamiltonian written down in terms of the Fermionic operators: 
\[
H=-\frac{1}{2}\left(\sum_{i=1}^{L-1}~a_i^{+}a_{i+1}+a_1^{+}a_L e^{i\pi(M-1)}
+h.c.\right),
\]
where $M$ is the number of particles. We assume for simplicity $L$- to be even and 
$M=L/2$ to be odd ($S^z=0$ for the ground state and $(M-1)$- even, we also 
assume $L$ to be even so that the ground state is not degenerate). Then the 
Hamiltonian is diagonalized with the help of the following Fourier transform 
in the different sectors:
\[
a_p^{+}=\frac{1}{\sqrt{L}}\sum_{x}e^{ikx}a_x^{+}, ~~~p=\frac{2\pi n}{L}, 
~~n\in Z, ~~(S^z=0),
\]
\[
c_q^{+}=\frac{1}{\sqrt{L}}\sum_{x}e^{iqx}a_x^{+}, ~~~q=\frac{2\pi (n+1/2)}{L}, 
~~n\in Z, ~~(S^z=-1).
\]
The formfactor of the operator $\sigma^{-}_L$ at the $L$-th site (which can be  
denoted also as $\sigma_0^{-}$) takes the form:
\be
\la\{q\}|\sigma^{-}_0|\{p\}\ra=\la0|\left(\prod_{i=1}^{M-1}c_{q_i}\right)
\frac{1}{\sqrt{L}}\sum_ka_k\left(\prod_{i=1}^M a_{p_i}^{+}\right)|0\ra,
\label{ac}
\ee
where the sets of the momenta $\{p\}= \{p_1,\ldots p_M\}$ 
and $\{q\}= \{q_1,\ldots q_{M-1}\}$ correspond to the eigenstates in the sectors with 
different number of particles (determined by the integers or half- integers numbers). 
The set $\{p\}$ corresponds to the ground state of the system while the set $\{q\}$ 
corresponds to the arbitrary excited state in the sector $S^z=-1$. 
One can represent the formfactor as a determinant using Wick's theorem and the 
following expression for the average:  
\be
\la0|c_q a_p^{+}|0\ra= \frac{2}{L}\frac{e^{i(p-q)}}{(1-e^{i(p-q)})}= 
\frac{i}{L}~\frac{e^{i(p-q)/2}}{\sin((p-q)/2)}.
\label{init}
\ee
Then we obtain the formfactor as a sum of determinants in the form 
\[
\la\{q\}|\sigma^{-}_0|\{p\}\ra= \frac{1}{\sqrt{L}} e^{-iq/2}
\sum_{m=1}^{M}(-1)^{m}e^{-ip_m/2}~\det_{ij}^{(M-1)}\left( 
\frac{i}{L}~\frac{1}{\sin((p_i^{(m)}-q_j)/2)}\right), 
\]
where the set of the momenta $\{p^{(m)}\}$ is obtained from the set $\{p\}$
by the exclusion of the single momenta $p_m$ and $q=\sum_{i=1}^{M-1}q_i$.  
Thus we obtain the following expression for the formfactor:
\[
\la\{q\}|\sigma^{-}_0|\{p\}\ra= 
\frac{1}{\sqrt{L}}\det_{ij}(M_{ij}(p,q)), ~~i,j=1,2,\ldots M, 
\]
where $M\times M$ matrix $M_{ij}$ equals: 
\[
M_{ij}=\frac{i}{L}~\frac{1}{\sin((p_i-q_j)/2)}, ~~j=1,\ldots M-1, 
\]
\be
M_{iM}=e^{-ip_i/2}, ~~ i=1,\ldots M.
\label{myform}
\ee
This determinant can be calculated in the following way. Introduce the 
new set of the momenta $\{q'\}= \{q_1,\ldots q_{M-1},q_M\}$ and take the limit
$q_M=iQ$, $Q\rightarrow\infty$. Then 
\[
\frac{i}{L}\frac{1}{\sin((p_i-iQ)/2)}\rightarrow -\frac{2}{L}e^{-Q}e^{-ip_i/2}, 
\]
and the formfactor is represented as a Cauchy determinant: 
\[
\det_{ij}(M_{ij}(p,q))=\frac{L}{2i}e^{Q}\det_{ij}
\left( \frac{i}{L}\frac{1}{\sin((p_i-q'_{j})/2)} \right). 
\]
Using the well known formula for this determinant and taking the limit 
$Q\rightarrow\infty$, we get the following formula for the formfactor: 
\be
\psi(\{q\})=
\frac{1}{\sqrt{L}}
\left(\frac{i}{L}\right)^{M-1} \left( e^{i\sum_{i=1}^{M-1}q_i} \right)
\frac{\prod_{i<j}\sin((p_i-p_j)/2)\prod_{i<j}\sin((q_i-q_j)/2)}{\prod_{i,j}
\sin((p_i-q_j)/2)}, 
\label{ff}
\ee
where the relation $\sin((p_i-iQ)/2)\rightarrow\frac{1}{2i}e^{ip_i}e^{Q}$ was used 
and the products are over the initial sets of the momenta $\{p\}$ and $\{q\}$. 
Let us stress that the expression (\ref{ff}) for the formfactor is valid not only 
for the ground state configuration of the momenta $\{q\}$ but for an arbitrary 
excited state characterized by the momenta $q_1,\ldots q_{M-1}$. 
The advantage of the expression (\ref{ff}) for the formfactors in comparison with 
the determinant expressions is that using the expression as a product one can hope 
to extract the behaviour of the formfactors on the momenta $q$ for the low-energy 
excited states (for example, the one-particle - one-hole excitations) and thus 
extract the asymptotic behaviour of the correlator.   
Note, that the expression (\ref{ff}) is valid for an arbitrary filling fraction 
(arbitrary $M$) or for the XX spin chain in the magnetic field.

Let us comment on the expressions for the formfactors obtained in ref.\cite{Colomo}. 
One can use the explicit expressions for the wavefunctions for the hard-core bosons 
in the coordinate space 
\[
\Psi(x_1,\ldots x_M)= \frac{1}{M!}\prod_{i<j}\epsilon(x_i-x_j)\sum_P (-1)^P
e^{i\sum_{i=1}^M k_{Pi}x_i}, 
\]
where $x_i$ is the coordinates of spins, $\epsilon(x)$ - is the sign function 
and $P$ - is the permutation of $M$ particles, $P\in S_M$,  
obtained as a limit of the Bethe wave function for the XXZ - chain and 
using the formula 
\[
\sum_x \epsilon(x) e^{i(p-q)x}=i\ctg((p-q)/2)
\]
to derive the following formula for the formfactor for the same 
sets of the momenta $\{p\}$ and $\{q\}$: 
\[
\la\{q\}|\sigma^{-}_0|\{p\}\ra= \det_{ij}(M_{ij}(p,q)), ~~i,j=1,2,\ldots M, 
\]
where $M\times M$ matrix $M_{ij}$ equals: 
\[
M_{ij}=\frac{1}{L}\ctg\left(\frac{p_i-q_j}{2}\right), ~~j=1,\ldots M-1, 
\]
\be
M_{iM}=1, ~~ i=1,\ldots M.
\label{ctg}
\ee
Let us show the equivalence of this expression with the expression 
given by eq.(\ref{myform}). The expression (\ref{init}) can also be represented as
\[
\la0|c_q a_p^{+}|0\ra= \frac{i}{L}\left(\ctg((p-q)/2)+i\right). 
\]
Then we obtain from (\ref{ac}) the determinant of the matrix of the same form as 
(\ref{ctg}) with the matrix elements $M_{ij}$ for $j\neq M$ replaced by 
$-i\ctg((p_i-q_j)/2)+1$. To show the equivalence of two expressions for the 
formfactor one can use the following theorem.  
Consider the determinant of the sum of two matrices 
$\det_{ij}(M_{ij}+a_{ij})$, where the second matrix $a_{ij}=c_j\phi_i$ -
is the matrix of rank $1$. Then we have: 
\[
\det_{ij}\left(M_{ij}+a_{ij}\right) = \det_{ij}(M_{ij}) + 
\sum_{k=1}^{M} c_k \det_{ij}(M_{ij}^{(k)}), 
\]
where the matrices $M_{ij}^{(k)}$ differ from the initial matrix $M_{ij}$ 
only by the substitution of its $k$-th column by $\phi_i$:
\[
M_{ij}^{(k)}=(1-\delta_{jk})M_{ij}+ \delta_{jk}\phi_i. 
\]
Applying this statement in the case of the matrix $a_{ij}=1$ ($c_j=1$, $\phi_i=1$), 
to the determinant obtained from (\ref{ac}) one readily get (apart from the irrelevant 
phase factor) the expression (\ref{ctg}). Thus the equivalence of the formfactor 
(\ref{ff}) with that obtained in \cite{Colomo} is shown.

  Using the expression for the formfactor (\ref{ff}), it is easy to obtain 
another expression for the formfactor corresponding to the ground-state 
configuration $\{p\}=\{p_1,\ldots p_M\}$, $p_i=2\pi/L(i-(M+1)/2)$, ($M$- is odd), 
and the ground state in the sector with $M-1$ particles 
$\{q_0\}=\{q_1,\ldots q_{M-1}\}$, $q^{(0)}_i=2\pi/L(i-M/2)$. 
Equivalently in eq.(\ref{myform}) one can take the shifted momenta 
\[
p_i=(2\pi/L)(i),~~i=1,\ldots M, ~~~~q^{(0)}_j=(2\pi/L)(j+1/2), ~~j=1,\ldots M-1. 
\]
and introduce an extra momentum $q_M=(2\pi/L)(M+1/2)$ into the products of 
eq.(\ref{ff}) in order to make the correspondence with the expressions for 
the building blocks for the correlator $R_N$ introduced in the next section 
more transparent. In fact, one can use the simple formulas for the product of 
sinuses \cite{Brychkov} 
\[
\prod_{k=1}^{M-1}\sin(\frac{\pi}{L}(k))=2^{(1-L)/2}\sqrt{L},  ~~~~~
\prod_{k=0}^{M-1}\sin(\frac{\pi}{L}(k+1/2))=2^{(1-L)/2},   
\]
where $M=L/2$ ($L$ is even) is taken into account, 
to obtain instead of (\ref{ff}) the product with equal 
numbers of variables $p_i$ and $q_j$: 
\be
\psi(\{q^{(0)}\})=\frac{1}{L}\left(\frac{i}{L}\right)^{M-1} 
\frac{\prod_{i<j}\sin((p_i-p_j)/2)\prod_{i<j}\sin((q_i-q_j)/2)}{\prod_{i,j}
\sin((p_i-q_j)/2)},
\label{prodform}
\ee
where the new sets of the momenta $\{p\}$ and $\{q\}$ are: 
\[
p_i=(2\pi/L)(i),~~~~q_i=(2\pi/L)(i+1/2),~~~~i=1,\ldots M. 
\]
In the equivalent form the expression (\ref{prodform}) can be again 
represented as a Cauchy determinant of the $M\times M$ matrix: 
\be 
\psi(\{q^{(0)}\})=\frac{1}{\sqrt{L}}\det_{ij}^{(M)}\left( 
\frac{i}{L}~\frac{1}{\sin(\frac{\pi}{L}(i-j-1/2))}\right),~~~~
M=\frac{L}{2}.  
\label{newform}
\ee 
This equation for the formfactor will be used in the next section 
to compare the constant $C_0$ for the asymptotics of the correlator 
$G(x)$ with the square of this formfactor (\ref{newform}) and obtain 
the new expression for the constant $C_0$ as a Cauchy determinant.  

Let us show that both expressions for the formfactors 
(\ref{myform}) and (\ref{ctg}) leads to the same expression for the 
correlator as an $M\times M$ determinant \cite{Colomo}, which in turn is 
equivalent to the exact expression for the correlator, first obtained in 
ref.\cite{LSM}.  
We consider the square of the formfator (\ref{ctg}) and sum over the 
intermediate states $\{q\}$ with the weight $e^{iqx}$, where 
$q=\sum_{i=1}^{M-1}q_i$. Since the formfactor is an antisymmetric function 
of the momenta $q_i$, the sum over each $q_i$ can be extended to the whole 
region $q\in(-\pi;\pi)$ independently of the other $q_i$.  
We also represent by the single determinant 
the product of the determinants of two matrices.  
One can use the following formula for the matrix 
elements (\ref{ctg}),  
\[
\ctg(\frac{p_i-q}{2})\ctg(\frac{p_j-q}{2})=
\ctg(\frac{p_i-p_j}{2})\left(\ctg(\frac{p_i-q}{2})-
\ctg(\frac{p_j-q}{2})\right)-1, 
\]
and the following Fourier transform, 
\[
\frac{1}{L}\sum_{q}e^{iqx}\ctg((q-p)/2)= i(1-\delta_{x,0})e^{ipx}. 
\]
Thus, using the formfactors in the form of the double sum of 
$(M-1)\times(M-1)$ determinants (with alternating signs and the sets of the 
momenta $\{p^{(n)}\}$ which is obtained from the set $\{p\}$ by omitting the 
momentum $p_n$) one easily obtains the following expression for the 
equal-time correlator as a difference of two $M\times M$ determinants 
depending on the distance $x$: 
\be
G(x)=\det_{ij}^{(M)}\left(M_{ij}+R_{ij}\right)-
\det_{ij}^{(M)}\left(M_{ij}\right), 
\label{Mcorrel}
\ee
where the matrices $M_{ij}$, $R_{ij}$ are given by the equations: 
\[
M_{ij}= \delta_{ij}(1-\frac{2x}{L})-(1-\delta_{ij})
\frac{2}{L}\frac{\sin((p_i-p_j)x/2)}{\tg((p_i-p_j)/2)}, ~~~~~
R_{ij}=\frac{1}{L}e^{i(p_i+p_j)x/2},         
\]
which was presented in ref.\cite{Colomo}. 
From the determinant expression for the correlator (\ref{Mcorrel}) 
one can obtain the exact expression for the correlator $G(x)$ on a finite 
lattice which is obtained in the next section with the help of the method 
proposed in ref.\cite{LSM}.

\vspace{0.2in}

{\bf 3. Exact correlation function: momentum distribution.} 

\vspace{0.1in}

Let us briefly review the exact calculation \cite{LSM} of the spin-spin 
equal-time correlation function (density matrix) for the XX- spin chain on 
finite lattice of the length $L$ \cite{O}:  
\[
G(x)= \la0|\sigma^{+}_{i+x}\sigma^{-}_{i}|0\ra.
\]
Using the Jordan-Wigner transformation relating spin operators to the 
Fermi operators ($a^{+}_i,~a_i$)  
$\sigma^{+}_{x}=\exp(i\pi\sum_{l<x}n_l)a^{+}_{x}$, the correlation function 
$G(x)$ can be represented as the following average for over the free-fermion 
ground state:
\[
G(x)= \la0|a^{+}_{x}e^{i\pi N(x)}a_{0}|0\ra, 
\]
where $N(x)=\sum_{i=1}^{x-1}n_i$. 
Introducing the operators, anticommuting at different sites,  
\[
A_i=a^{+}_i+a_i, ~~~~~B_i=a^{+}_i-a_i, ~~~~A_iB_i=e^{i\pi n_i}, 
\]
where $n_i=a^{+}_ia_i$ - is the fermion occupation number, with the following 
correlators with respect to the free-fermion vacuum, 
\[
\la0|B_iA_j|0\ra=2G_0(i-j),~~~\la0|A_iA_j|0\ra=0,~~~ \la0|B_iB_j|0\ra=0,
\]
where the free-fermion Green function on finite chain $G_0(x)$ is 
\[
G_0(x)=\la0|a^{+}_{i+x}a_{i}|0\ra=\frac{\sin(\pi x/2)}{L\sin(\pi x/L)}, 
\]
one obtains the following expression for the bosonic correlator:
\[
G(x)=\frac{1}{2}\la0|B_0(A_1B_1)(A_2B_2)\ldots(A_{x-1}B_{x-1})A_x|0\ra.
\]
Note that we assume the periodic boundary conditions for the initial spin 
operators, so that strictly speaking, the above formulas are valid only for 
the case when $L$ - is even (the ground state in not degenerate) and 
$M=L/2$ - is an odd integer (so that $L/4$ - is not an integer).  
That is easily seen from the boundary term 
$\sim\exp(i\pi(M-1))=+1$ when $M$- is odd, momenta of fermions - are integers 
(not half-integers) and their configuration is symmetric around zero. 
In that case the free- fermionic function $G_0(x)$ - is given exactly by the 
above formula.    
(as for the case $M$ - even, it is clear from the general grounds that in this case 
Eq.(9) will be modified by the terms of order $1/L$ for any $x$).   
Using Wick's theorem we obtain the following determinant of $x\times x$ matrix:
\[
G(x)=\det_{ij}(2G_0(i-j-1)), ~~~~i,j=1,\ldots x.
\]
Due to the form of this matrix ($G_0(l)=0$ for even $l$) this determinant 
can be simplified and the following formulas are obtained: 
\[
G(x)=\frac{1}{2}(R_N)^2,~(x=2N),~~~~G(x)=-\frac{1}{2}R_{N}R_{N+1},~(x=2N+1), 
\]
where we denote by $R_N$ the following determinant of the $N\times N$- matrix: 
\[
R_N=\det_{ij}\left((-1)^{i-j}2G_0(2i-2j-1)\right),  ~~~~i,j=1,\ldots N, 
\]
where $G_0(x)$ is the same Green function of free fermions as above. 
Since $R_N$ - is the Cauchy determinant one can obtain the following 
expression for it on the finite chain: 
\be
R_N= \left(\frac{2}{\pi}\right)^N
\prod_{k=1}^{N-1}\left(\frac{(\sin(\pi(2k)/L))^2}{\sin(\pi(2k+1)/L)
\sin(\pi(2k-1)/L)}\right)^{N-k}. 
\label{sinprod}
\ee
Note that we obtained the exact expression for the correlator on the 
finite lattice \cite{O}. From the expression (\ref{sinprod}) it is easy to 
obtain the correlator in the thermodynamic limit ($L\rightarrow\infty$)
which is given by the similar product. For finite chain it is easy 
to evaluate (\ref{sinprod}) numerically and compare the result with
the asymptotic (\ref{coas}) at $x>>1$ and $x\sim L$. 
At the distances $x\sim L$ the correction to the asymptotic formula 
(\ref{coas}) behaves like $\sim 1/L$.  
We find that the exact correlator coincides with 
the correlator given by (\ref{sinprod}) with very high accuracy 
up to the very small distances $x\sim 1$.

Using the expression (\ref{sinprod}) one can get the new expression for 
the constant $C_0$ for the leading asympotics for the corellator $G(x)$ 
(\ref{coas}). For convenience one can redefine the spin operators according to 
$\s_{x}^{\pm}\to(-1)^{x}\s_{x}^{\pm}$, which is equivalent to the change of 
the sign for the Hamiltonian of the XX chain. That leads to the shift of the 
singularity in momentum distribution from the values $q=\pm\pi$ to $q=0$, 
which will be convenient later in this section for evaluating the momentum 
distribution. Then the correlator (\ref{coas}) takes the form:    
\[
G(x)=C_0\frac{1}{(L\sin(\pi x/L))^{\a}},  
\] 
($\a=1/2$). According to this formula the correlator $G(x)$ at the distance 
$x=L/2$ equals $G(L/2)=C_0/\sqrt{L}$ with the corrections of higher order in $1/L$.  
On the other hand, since our expressions for the correlatar are obtained for 
$M=L/2$ - odd, we have $G(L/2)={1\over 2}R_{\frac{M-1}{2}}R_{\frac{M+1}{2}}$, 
where $R_N$ is given by eq.(\ref{sinprod}), or, equivalently, 
\[
R_N=\det_{ij}\left(\frac{2}{L\sin\left((2\pi/L)(i-j-1/2)\right)}\right), 
~~~~i,j=1,\ldots N. 
\]
Clearly, this determinant can also be represented as the product of sinuses. 
First, this relation can be used to obtain the new expression for the 
constant $C_0$. Second, one can use the relation $G(L/2)=C_0/\sqrt{L}$ to 
predict the asymptotic behaviour of the last dererminant $R_N$ at large $L$ 
(or $M$). Next, one can compare $R_N$ with the expression (\ref{newform}) 
for the formfactor for the intermediate state corresponding to the ground 
state in the sector with $M-1$ particles. Note that there is an obvious   
similarity between the expressions for two different quantities. 
In particular using these arguments one can predict the asymptotic 
behaviour of the formfactor $\psi(\{q^{(0)}\})$ at large $L$ including 
the constant proportional to $C_0$ which is interesting by itself 
since in general it could allow to relate the constant $C_0$ to the square of 
the lowest formfactor.

In fact, neglecting the terms of order $1/L$, one gets the relation 
\[
\left(R_{M/2}\right)^2=\frac{C_0\sqrt{2}}{\sqrt{M}}, 
\]
where $R_{M/2}$ equals to the determinant which is expressed through $M$ as 
\be
R_{M/2}=
\det_{ij}\left(\frac{1}{M\sin\left((\pi/M)(i-j-1/2)\right)}\right), 
~~~~i,j=1,\ldots M/2. 
\label{rm}
\ee
Taking this expression into account one can find both the scaling 
behaviour (the behaviour in a power of $L$) and the constant for 
the determinant in eq.(\ref{newform}) determining the the square of the 
lowest formfactor $|\psi(\{q^{(0)}\})|^2$. In fact, using the scaling 
arguments, i.e. neglecting the corrections of order $1/L$, comparing 
the expression (\ref{rm}) with the determinant (\ref{newform}), 
one finds 
\be
|\psi(\{q^{(0)}\})|^2=\frac{C_{0}\sqrt{2}}{\sqrt{L}}. 
\label{lowest}
\ee
Thus we found the ``lowest'' formfactor i.e. the single formfactor, 
corresponding to the ground state in the complete set of the intermediate 
states. 
This formfactor can be compared with the Fourier transform of the 
correlator $G(x)$ at zero momentum, $\sum_{x=1}^{L}G(x)$, which gives the 
sum of the squares of the formfactors for all intermediate states with zero 
total momentum (including the ground state formfactor (\ref{lowest})). 
First, one observes that in order to calculate this quantity it is 
sufficient to use the leading order expression (\ref{coas}) for the total 
correlator $G(x)$. In fact, it is easily seen that the sum for the 
subleading terms is suppressed by the powers of $L$. 
Second, one can see that to calculate the sum $\sum_{x=1}^{L}G(x)$ it is 
sufficient to replace the sum by the corresponding integral since the 
corrections are again of order $\sim 1/L$. Denoting the corresponding 
sum of the formfactors as $|\psi_0|^2=(1/L)\sum_{x=1}^{L}G(x)$ we obtain 
\be
|\psi_0|^2=\frac{C_1}{\sqrt{L}},~~~~
C_1=C_{0}\int_{0}^{1}dy\frac{1}{(\sin(\pi y))^{1/2}},  
\label{all}
\ee
which clearly has the same scaling behaviour as eq.(\ref{lowest}). 
The constant $C_1$ in the last equation should be compared with 
the constant $\sqrt{2}C_0$ in (\ref{lowest}). Calculating the integral, 
one finds the coefficient 
$C_1=\sqrt{\pi}\Gamma(1/4)/\Gamma(3/4)C_0=1.6725..C_0$ 
which is to be compared with $\sqrt{2}C_0$. 
One finds the striking coincidence  of two different 
values, which can be explained only by the extreme smallness of 
the contributions of the excited states formfactors to the sum 
(\ref{all}) in comparison with the ``lowest `` formfactor 
(\ref{lowest}). 
In our opinion this result is of interest in connection with the 
study of the correlators for the general case of the XXZ spin chain. 
The smallness of the higher states contributions in (\ref{all}) 
can also be seen from the representation (\ref{ff}). For example, 
consider the configuration $\{q\}$ with $q=0$ with $q_1$ and 
$q_{M-1}$ shifted by one step to the left and to the right 
respectively. One can easily calculate the value of (\ref{ff}) 
which is lowered by the factor $1/4$ with respect to the ground state. 
Taking the square of this value one gets the factor $1/16$ which is 
close to the relative difference between (\ref{lowest}) and 
(\ref{all}).  

Let us turn to the study of the momentum distribution for the XX  
spin chain which turns out to be completely different from that for 
the system of fermions (Luttinger model) (\ref{nkf}). 
The results obtained are equally well applicable to the general case 
of the XXZ chain provided the constant for the asymptotics (\ref{coas}) 
for XXZ case is known \cite{Luk}. 
Turning back to the sum of the formfactors $|\psi_0|^2$ (\ref{all}) one 
can see that this quantity is nothing else but the probability $n(q=0)$ 
to have the total momentum $q=0$ which is defined by the equations: 
\be
n(q)=\frac{1}{L}\sum_{x}e^{-iqx}G(x), ~~~G(x)=\sum_{q}e^{iqx}n(q),~~~
n(q)=\la\s_q^{+}\s_q^{-}\ra,  
\label{mo}
\ee
where the sum over $q=2\pi n/L$, $n\in Z$ is extended over the interval 
$q\in(-\pi,\pi)$, $\s_q^{+}$ and $\s_q^{-}$ are the Fourier transform 
of the initial spin operators and for the half-filling the normalization 
condition is $\sum_{q}n(q)=1/2$. The value of $n(0)=C_1/\sqrt{L}$ was 
already calculated above (\ref{all}) (the useful formulas which allow 
one to perform the numerical estimates are presented in the Appendix,  
note that the value of $n(0)$ can be easily obtained for the XXZ chain).   
Let us stress once more that the value $|\psi_0|^2=n(0)$ 
($=C_1/\sqrt{L}$) is an exact value of $n(0)$ - the sum of the 
formfactors with $q=0$ in the limit of the large chain $L>>1$. 
Next, from (\ref{coas}) one can calculate $n(q)$ for the XXZ chain 
for the sufficiently small $q$ in the form: 
\be
n(q)=\frac{C_0}{L^{\a}}\sin(\frac{\pi\a}{2})
\frac{2^{\a}\Gamma(2-\a)}{\pi(1-\a)}
\frac{\Gamma(n+\a/2)}{\Gamma(n+1-\a/2)},~~~~q=\frac{2\pi n}{L}. 
\label{nnq}
\ee
Evidently, at $q=0$ one obtains the constant $C_1$ calculated above. 
This formula is correct provided the sum over $x$ in eq.(\ref{mo}) 
can be replaced by the integral i.e. for $n<<L$ since 
at larger values of $q$ the oscillatory behavior should strongly 
suppress $n(q)$ up to the value $\sim1/L$ at $q\sim\pm\pi$. 
The behaviour in $n$ can be found using the formula for the ratio of 
two Euler's gamma- functions in the equation (\ref{nnq}).   
For the XX chain it predicts the behaviour of order $1/\sqrt{L}\sqrt{n}$ 
at $1<<n<<L$ where $n=Lq/2\pi$.  Clearly, at $n\sim1$ or $q\sim2\pi/L$ 
this formula is not valid, so at $n\sim1$ one should use the initial 
formula (\ref{nnq}). 
In the general case of the XXZ chain the 
equation (\ref{nnq}) gives $1/L^{\a}n^{1-\a}$. 
The constant in front of this asymptotics for $n(q)$ is not equal 
to the constant $C_1$ calculated above. However, it can be easily 
found for the XX chain from eq.(\ref{nnq}) as 
$C_2=(2/\pi)\Gamma(3/2)C_0=(1\sqrt{\pi})C_0=0.564..C_0$.  
As in the case of the Luttinger model, 
the contribution of the subleading terms in $G(x)$ are suppressed 
by a powers of $1/L$ in the region, where the asymptotics 
\be
 n(q)=\frac{C_2(\a)}{L^{\a}}\frac{1}{n^{1-\a}}=
\frac{C'_2(\a)}{L}\frac{1}{q^{1-\a}}
\label{nn}
\ee
is valid i.e. in the region where the value of $n(q)$ is parametrically 
larger than $1/L$. In this sense the momentum distribution given by the 
equation (\ref{nn}) is exact. The constants $C_2$, $C'_2$ for the XXZ 
chain equal 
\[
C_2(\a)=\sin(\frac{\pi\a}{2})\frac{2^{\a}\Gamma(2-\a)}{\pi(1-\a)}C_0, 
\]
and $C'_2(\a)=C_2(\a)(2\pi)^{1-\a}$. For the XX spin chain we get 
$C_2=0.564..C_0$, $C'_2=1.410..C_0$. 
There is no reason to expect that the ``singular'' part of 
the function $n(q)$, $n(q)=C'_2/Lq^{1-\a}$ (\ref{nn}) saturates the 
sum rule 
\[
\sum_{q=-\pi}^{\pi}n(q)=\frac{1}{2}, 
\]
since in general the subleading terms give the contribution of order 
$L(1/L)\sim1$. However one can calculate the contribution of the 
function (\ref{nnq}), $\sum_{n=0}^{L}n(q)$ taking into account the 
known value of the constant $C_0=2\sqrt{\pi}(0.147088..)=0.5214..$ 
(see Appendix). 
In fact, calculating the sum of (\ref{nnq}) we obtain 
$0.588..$, which is again close to the total value $1/2$. 
This fact is in agreement with the small value of the subleading term 
in the correlator $G(x)$ (see Appendix) and the observation \cite{O} 
that the exact correlator is extremely close to its asymptotic value 
(\ref{coas}) at an arbitrary distance $x=1,\ldots L$. 
The same estimates can be performed for the case of the XXZ spin chain 
which leads to the similar results and in fact can be used to obtain 
the sufficiently accurate estimate for the constant $C_0$ \cite{Luk}. 
Note, that exactly the same relation for the spinless fermion model 
(Luttinger model) between the constant $C(p_F,\Delta)$ in the equation 
(\ref{sinas}) and the constant $C$ in the equation (\ref{nkf}) 
takes place: one should simply replace the exponent $\a$ by $\a(\l)$. 
In the context of the possible application for the XXZ model, it would 
be interesting to predict the asymptotic behaviour of the momentum 
distribution (\ref{nn}) and the function (\ref{coas}) from 
the beheviour of the formfactors (\ref{ff}) for the low-lying 
excitations (small total momentum $q$).  
Note also that the calculation of the constant $C_0$ in the model of 
spinless fermions (see eq.(\ref{sinas}) as well as in the other models 
(for example, for the XXZ spin chain in the magnetic field) 
remains an open problem.

\vspace{0.2in}

{\bf Conclusion.}

\vspace{0.1in}

In conclusion, we presented the new expressions for the formfactors of 
local operators for the XX - quantum spin chain as a Cauchy determinants. 
Using the functional form of the correlator at large distances we proposed  
the new expression for the constant for the asymptotics of the correlator 
as a Cauchy determinant. Using the scaling arguments, the value of the 
``lowest'' formfactor for the XX chain was found. 
Using the functional form of the correlator the momentum distribution for 
the XXZ spin chain is evaluated.  
The universal character of the connection of 
the momentum distribution singularity with the asymptotics of the correlators 
and the constants in front of the asymptotics both for the XXZ spin chain and 
the Luttinger model (spinless fermion model) was pointed out. 
It is possible that the formfactor approach 
can be useful for the calculation of the constant $C_0$ in the general case of 
the XXZ - spin model and the other models solvable by the algebraic Bethe 
ansatz method.

\vspace{0.3in}

{\bf Appendix.} 

\vspace{0.1in}

For completeness we present here  
the formulas for the integrals required for the numerical estimate 
of the constant $C_1$ in the text and present some of the results   
obtained previously for the constant $C_0$.

First, the integrals one can use are \cite{Brychkov}: 
\[
\int_{0}^{\pi}\frac{1}{(\sin(y))^{\a}}dy=2^{-\a}
B\left(\frac{1-\a}{2};\frac{1-\a}{2}\right)=
\sqrt{\pi}\Gamma\left(\frac{1-\a}{2}\right)
\left(\Gamma(1-\frac{\a}{2})\right)^{-1},    
\]
where $B(x;y)=\Gamma(x)\Gamma(y)/\Gamma(x+y)$, 
$\Gamma(1/2)=\sqrt{\pi}$, $\Gamma(1/4)=3.625600..$, $\Gamma(3/4)=1.225417..$,  
which is sufficient to estimate the constant 
$C_1=\sqrt{\pi}(\Gamma(1/4)/\Gamma(3/4))C_0$.  
To obtain the formula for the momentum distribution presented in the 
text one can use the following integral: 
\[
\int_{0}^{\pi}e^{iqy}\frac{1}{(\sin(y))^{1-\nu}}dy= 
\frac{\pi e^{iq\pi/2}}{2^{\nu-1}\nu 
B\left(\frac{\nu+q+1}{2};\frac{\nu-q+1}{2}\right)} 
\]
and take into account the well known relation 
$\Gamma(1-x)\Gamma(x)=\pi/\sin(\pi x)$. 
The correspondence between this two formulas can be established with the 
help of the relation 
$\Gamma(2z)=(2^{2z-1}\pi^{-1/2})\Gamma(z)\Gamma(z+1/2)$. 
To obtain the asymptotic formula (\ref{nn}) one should use the well known 
expression for the asymptotics of $\Gamma(z)$ at large $z$. 

For completeness let us present here some results obtained previously 
for the correlator $G(x)$ in the thermodynamic limit. 
Using the expression (\ref{sinprod}) in the 
thermodinamic limit we calculate the product: 
\[
R_N= \left(\frac{2}{\pi}\right)^N
\prod_{k=1}^{N-1}\left(\frac{(2k)^2}{(2k+1)(2k-1)}\right)^{N-k}. 
\]
Considering the logarithm of $R_N$ after some algebra  
one can finally obtain the result:
\be
\ln(R_N)= -\frac{1}{4}\ln(N)+\frac{1}{4}\int_0^{\infty}\frac{dt}{t}\left(
e^{-4t}-\frac{1}{(\ch(t))^2}\right)-\frac{1}{64~N^2}, 
\label{int}
\ee
where the omitted terms are of order $\sim 1/N^4$. 
The above 
expression coincides with the formula proposed in ref.\cite{Luk} in the 
particular case of the XX - chain. 
The last term in (\ref{int}) gives the following coefficient for the 
next-to-leading asymptotics for the correlator: 
\be
G(x)\simeq \frac{C_0}{\sqrt{\pi}}\left((-1)^x\frac{1}{x^{1/2}}
-\frac{1}{8}\frac{1}{x^{5/2}}\right),
\label{asympt}
\ee
where the constant $C_0$ is defined in (\ref{coas}). 
Remarkably, the equation (\ref{asympt}) shows that the exact correlator 
$G(x)$ coincides with the leading asymptotic result (\ref{coas}) even 
in the region $x\sim1$ with the sufficiently high accuracy. 
The value of the constant corresponing to the subleading term in eq.(\ref{asympt})
was first obtained by McCoy \cite{McCoy} using the method \cite{Wu} 
with the help of the asympotics of the Barnes $G$- function \cite{Barnes}, 
defined by $G(z+1)=\G(z)G(z)$, $G(1)=1$.  
The general expression for the Cauchy determinant is: 
\be
\det_{ij}^{(N)}\left(\frac{1}{i-j+z}\right)= 
\frac{(G(N))^2G(1+z)G(1-z)}{G(1+z+N)G(1-z+N)}
\left(\frac{\pi}{\sin(\pi z)}\right)(-1)^{\frac{N(N-1)}{2}}. 
\label{z}
\ee
In particular, the product $R_N$ can be represented as 
\[
  R_N=(G(1/2))^2\frac{(G(N+1))^2}{G(N+1/2)G(N+3/2)}.  
\]
Using the asymtotics of the function $G(N)$ at large $N$, 
\[   
     G(N)= \frac{1}{12}-\ln A - \frac{1}{2}(\ln2\pi)N + \left(\frac{1}{2}N^2
- \frac{1}{12}\right)\ln N - \frac{3}{4}N^2 +O\left(\frac{1}{N^2}\right), 
\]
where $A$ is the Glaisher constant (see below), one can obtain 
the result (\ref{asympt}) with the constant 
$(C_0)^{1/2}=\pi^{1/4}2^{1/12}e^{1/4}A^{-3}$. 
One can evaluate the integral in (\ref{int}) to get the asymptotic 
\[
\ln(R_N)= -\frac{1}{4}\ln(N)+ \left(\frac{\ln2}{12}+3\zeta'(-1)\right),~~~ 
\]
which is equivalent to the estimate 
\[
R_N=\left(N^{-1/4}\right)\left(2^{1/12}e^{1/4}A^{-3}\right), ~~~~
A=e^{1/12-\zeta'(-1)}=1.282427 \ldots. 
\]
This result agrees with the result obtained by Wu \cite{Wu}, using the 
expression of the product through the Barnes G- functions \cite{Barnes}.   
Thus for the constant before the asymptotic (\ref{coas}) the known value 
$C_0/2\sqrt{\pi}=0.147088\ldots$ is obtained.

\end{document}